# EASTER EGGS, MYTHS AND JOKES IN FAMOUS PHYSICS BOOKS AND PAPERS

**Lorenzo Fortunato**

**Dip. Fisica & Astronomia "G.Galilei" - Univ. Padova (Italy) & I.N.F.N.**

**Apr., 1st 2017**

Scientists, and physicists in particular, are a kind of revolutionary people, they try to innovate, to reinvent, to turn the cards on the table and to investigate what happens when you break the rules. Often this attitude manifests itself in the realm of social conventions that the scientists smash just for fun or to see what happens in a sort of social experiment, in which, to say the truth, they themselves are the disturbing element. Everybody knows about the unpaired socks of Albert Einstein or about Richard Feynman playing the bongos in Brazil. This sort of devilish pranking mood is a common distinctive trait of true genius that many non-genial scientists pretend to emulate without success. In addition, even the stiffest and most formal professors have that sort of childish curiosity that pushes them to dedicate their life to science as if it was a funny game, to be played for the mere sake of playing. And as junkies addicted to game, they never really stop playing.

I believe these are two reasons that can help explaining why some of the greatest physicists sometimes have concealed *Easter eggs* in their manuscripts, i.e. they have hidden in their books and publications a totally absurd sentence, a gigantic joke, an everlasting reminder of the fallacy of our *streben* to keep an unstable equilibrium between seriousness and folly, or more probably a sign of their innate consciousness that *'Semel in anno licet insanire'*, that even to aspiring Nobel winners, once a year it is allowed to get crazy, as the Ancient Romans used to put it. Probably, sometimes they just wanted to scoff the editors, thinking that at some point a bored reviewer would inquire about the eccentric statements, stopping the printing of the book and demanding for polishing of the text.

I will report below on a few examples of raving and insane (or maybe utterly genial) sentences that can be found in famous and otherwise admirable books of physics, because I genuinely believe it is amusing.

### 1. Familiar Chinese characters

Probably the best known example is found in the introduction of the 1969 bible of nuclear structure Vol. 1, the famous book by Bohr & Mottelson [1]. Here the two Nobel Prize winners make use of a self-explaining and user-friendly type of labeling (see picture below). And they use it throughout the book to simplify the task of the readers.

> The division into text, illustrative examples, and appendixes is clearly indicated by the typography. To further help the reader, each page has been labeled by one of the familiar Chinese characters 文 (wen = text), 圖 (t'u = illustration), and 附 (f'u = appendix), which so graphically convey the distinction between the three divisions.

It is true that, statistically, they have chosen the more common language on this planet, and probably one of the most concise, but the tone of the sentence is self-explaining. I'm not expert on oriental languages and I can't tell if it is Mandarin or tangerine.

**2. Marilyn Monroe Equations**

At the page 159 of the celebrated book "The Harmonic Oscillator in Modern Physics" by Moshinsky and Smirnov (Harwood Academic Publisher, vol. 9 of Series Contemporary Concepts in Physics), amidst a bunch of hard quantum mechanics and group theory, all of a sudden one can read about the sex appeal of Marilyn Monroe that is, beyond any doubt, just like a proven mathematical theorem:

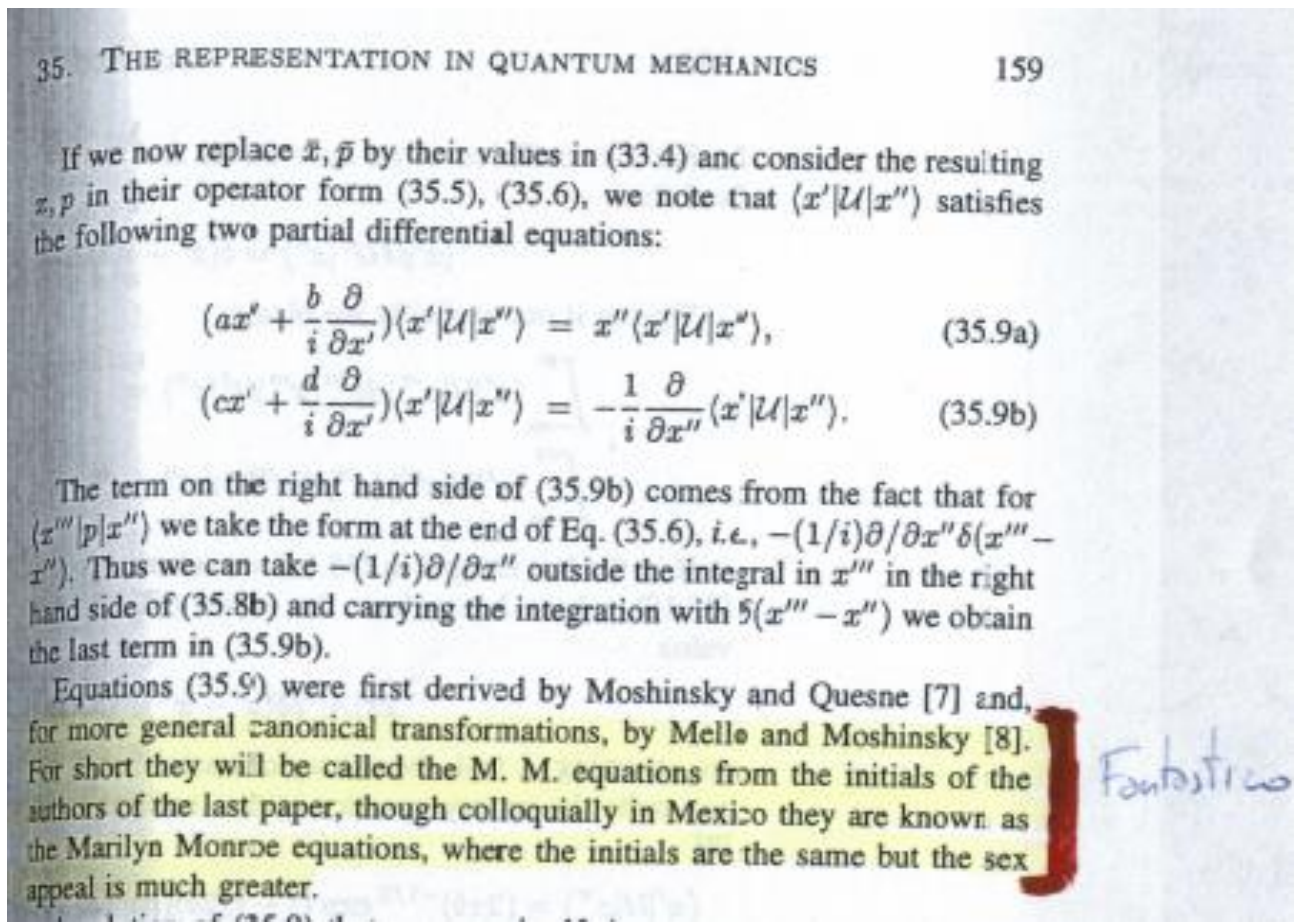

The yellow marked text (mine the colouring and comment in Italian) hits the unprepared reader as a punch on the teeth. I have carefully scanned the book looking for other concessions to common sense, but I haven't found any, they must have finished the tequila/vodka.

I must add, as a personal comment, that there is a sense of reward when you are digging in a book like this: you are working your way through equations that are heavy and hard as rocks and all of a sudden, boom, there you find a gold nugget, that is good for a thousand anecdotes at the coffee machine. I treat it and present it as if it was a personal discovery, like a topaz crystal found scraping the soil barehand.

## 3. "Pretty mathematics"

This is the title that Paul Adrien Maurice Dirac (a.k.a. PAM Dirac) gave to his contribution to the symposium held in his honour at the Loyola University, New Orleans, in May 1981, then published in the International Journal of Theoretical Physics, Vol. 21, Nos.8/9 (1982). The concise abstract, reported below, is acceptable only if you recall the fact that the 80 year old man at that time was a Nobel Prize winner from already half a century and both the age and the prize gave him enough credit and freedom to talk about whatever he wanted in the way he wanted. It reads:

*"Mathematical beauty is important in particle theories, two of which are described."*

Irritating and provocative, but dispassionate: his main point is that, if one is lucky enough, invoking aesthetic principles in mathematical theories might lead to insightful discoveries in those branches of physics that make use of these theories and this has sometimes been true, like in the case of the discovery of the positron, and sometimes not, like in the case of magnetic monopoles. It is remarkable that such a weak philosophical working principle has led to such epoch-making discoveries!

This concludes this brief survey of Easter Eggs in publications by famous physicists. I hope you enjoyed the reading and you are now possessed by that sort of gold fever that I was mentioning above. By the way, please have a look at the publication date. Appropriate isn't it?